\documentclass{article}

\usepackage{amsmath,amsfonts,amssymb,theorem}
\usepackage{graphics} 
\usepackage{psfrag}

\mathchardef\ordinarycolon\mathcode`\:     
\mathcode`\:=\string"8000
\def\vcentcolon{\mathrel{\mathop\ordinarycolon}} \begingroup
\catcode`\:=\active \lowercase{\endgroup \let :\vcentcolon }

\newcommand{\Gal}{\mathrm{Gal}}  
\newcommand{\Q}{{\mathbb Q}}
\newcommand{\Z}{{\mathbb Z}}  

\newcommand{\C}{{\mathbb C}} 
\newcommand{\F}{{\mathbb F}}
\newcommand{\Tr}{\mathrm{Tr}} 
\newcommand{\Prob}{\mathrm{Prob}}
\newtheorem{theorem}{Theorem}
\newtheorem{corollary}{Corollary}
\newtheorem{definition}{Definition}
\newtheorem{lemma}{Lemma}
\newtheorem{example}{Example}
\newtheorem{fact}{Fact}
{\theorembodyfont{\rmfamily} \newtheorem{algorithm}{Algorithm}}
\newenvironment{proof}{\noindent{\bf Proof: }} {\hfill\rule{2mm}{2mm}}
\newcommand{\ket}[1]{|#1\rangle} 

\newcommand{\smfrac}[2]{\mbox{$\frac{#1}{#2}$}}
\renewcommand{\root}[2]{\zeta_{#1}^{#2}} 
\newcommand{\e}{\mathrm{e}}
\renewcommand{\i}{\mathrm{i}}
\newcommand{\ox}{\otimes}

\title{{Efficient Quantum Algorithms for Estimating Gauss Sums}}
\author{Wim van Dam\footnote{HP Labs Palo Alto;  Mathematical Sciences
Research Institute, Berkeley; Computer Science Division, University of
California, Berkeley.  Supported by an HP-MSRI postdoctoral
fellowship.  Email: \texttt{vandam@cs.berkeley.edu} }\and Gadiel
Seroussi\footnote{HP Labs Palo Alto.  Email:
\texttt{seroussi@hpl.hp.com}}}
\begin{document}

\maketitle

\begin{abstract}
We present an efficient quantum algorithm for estimating Gauss sums
over finite fields and finite rings. This is a natural problem as the
description of a Gauss sum can be done without reference to a black
box function.  With a reduction from the discrete logarithm problem to
Gauss sum estimation we also give evidence that this problem is hard
for classical algorithms.  The workings of the quantum algorithm rely
on the interaction between the additive characters of the Fourier
transform and the multiplicative characters of the Gauss sum.
\end{abstract}

\section{Introduction}
Let $\chi:R\rightarrow\C$ be a multiplicative character and
$e:R\rightarrow\C$ an additive character over a finite ring $R$.  The
\emph{Gau{ss} sum} $G$ of this triplet $(R,\chi,e)$ is the inner
product between $\chi$ and $e$, that is: $G(R,\chi,e):=\sum_{x\in
  R}{\chi(x)e(x)}$.  Gau{ss} sums are useful on many fronts for the
analysis of finite fields $R=\F_{p^r}$ and rings $R=\Z/n\Z$.  In
combination with the closely related \emph{Jacobi sums}, they have
been used to prove theorems about Diophantine equations, difference
sets, primality testing, et cetera.  
One can view Gauss sums as the finite versions of 
the gamma function $\Gamma(s):=\int_0^\infty x^{s-1}\e^{-x}dx$.
See Brendt et al.\ 
\cite{gausssums} for a book entirely devoted to these topics.

The theory of quantum computation investigates if the laws of quantum
physics allow us to process information in a more efficient way than
is possible by the classical, Turing machine model of computation.
Strong support for the claim that quantum computers are indeed more
powerful than classical ones was given in 1994 by Peter Shor who
proved the existence of efficient quantum algorithms for factoring and
the discrete logarithm problem \cite{shor}.  More recently, Hallgren
showed that also Pell's equation can be solved in polynomial time on a
quantum computer \cite{pell}.  The common ingredient of these (and
other) quantum algorithms is the use of quantum Fourier transform to
extract the periodicity of an unknown function in time logarithmic in
the size of the domain.  See the book by Nielsen and Chuang for a
thorough introduction to this field \cite{nielsenchuang}.

In this article we describe a quantum algorithm that, given the
specification of the characters $\chi$ and $e$, efficiently
approximates the corresponding Gau{ss} sum, for $R$ a finite field
$\F_{p^r}$ or a `$\bmod{n}$' ring $\Z/n\Z$.  Because determining the
norm $|G|$ of a Gau{ss} sum is relatively straightforward, our
algorithm focuses on estimating the angle $\gamma\bmod{2\pi}$ in the
equation $G=|G|\cdot\e^{\i\gamma}$.  We describe a quantum transform
that induces this angle as a relative phase by a mapping
$\ket{0}+\ket{1}\mapsto\ket{0}+\e^{\i\gamma}\ket{1}$.  Because this
transformation can be implemented efficiently, we can sample the
output state $O(\smfrac{1}{\varepsilon})$ times to get an estimation
$\tilde{\gamma}$ of the angle $\gamma$ with expected error
$\varepsilon$.  The time complexity of this algorithm is 
$O(\smfrac{1}{\varepsilon}\cdot\mathrm{polylog}|R|)$.
Using a reduction from the discrete log problem to the
approximation of Gau{ss} sums, we provide evidence that this is a hard
task on a classical computer.  A discussion on the merits of Gau{ss}
sum estimation is included at the end of the article.

Section~\ref{sec:defs} gives the definitions and known results 
that we will use for the estimation of Gauss sums over finite fields.
The basic quantum procedures that we use for our algorithm
are defined in Section~\ref{sec:quant};  
the algorithm itself is described in Section~\ref{sec:alg}.
Next, we discuss the possibility of estimating Gauss sums with 
classical algorithms.  The relationship between this problem and 
the discrete logarithm problem, Galois automorphisms and random 
walks is explained in Section~\ref{sec:class}.
Section~\ref{sec:rings} gives some background that is necessary 
to define the Gauss sum problem for finite rings. 
A quantum algorithm for this problem is given in Section~\ref{sec:alg_ring}.
The final Section~\ref{sec:concl} discusses the connection 
between the presented algorithm for Gauss sum estimation and other quantum
algorithms.  Also the relative hardness of the problem with respect to
other known problems is addressed.
Throughout the article, results that are already known are indicated
as `facts'.

\section{Gauss Sums over Finite Fields}\label{sec:defs}
\subsection{Definitions and Notation: Gau{ss} Sums over Finite Fields}
Let $\root{p}{}$ denote the $p$th  root of unity:  $\root{p}{} :=
\e^{2\pi \i/p}$.  The \emph{trace} of an element $x$ of the finite
field $\F_{p^r}$ over $\F_p$ is $\Tr(x) := \sum_{j=0}^{r-1}{x^{p^j}}$.  It can be
shown that for every $x\in\F_{p^r}$, its trace is  an element of the
base-field: $\Tr(x)\in\F_p$.  For any $\beta\in\F_{p^r}$ we also have
the  related functions  $x\mapsto \Tr(\beta x)$.  These trace
functions are all the  linear functions $\F_{p^r}\rightarrow \F_p$
(note that $\beta = 0$ gives the trivial function $0$).  When we write
$\root{p}{\Tr(x)}$ we interpret the value $\Tr(x)$ as an element of
the set $\{0,1,\dots,p-1\} \subset \Z$.  For $\beta\in\F_{p^r}$, 
the functions  $e_\beta(x) := \root{p}{\Tr(\beta x)}$
describe all possible  additive characters $\F_{p^r}\rightarrow \C$.

Let $g$ be a primitive element of $\F_{p^r}$, i.e. 
the multiplicative group $\langle g \rangle$ generated by $g$ 
equals $\F^*_{p^r}$.
For each $0 \leq \alpha \leq p^r-2$, the
function  $\chi(g^j) := \root{p^r-1}{\alpha j}$ (complemented with
$\chi(0) := 0$) is a multiplicative character  $\F_{p^r} \rightarrow
\C$. Also, every multiplicative character can be written as such a
function.   For a non-zero $x\in\F_{p^r}^*$, the discrete logarithm
with  respect to $g$ is defined by $\log_g(g^j):=j\bmod{p^r-1}$.
Hence, every multiplicative character can be expressed by $\chi(x) :=
\root{p^r-1}{\alpha\log_g(x)}$ for $x\neq 0$ and $\chi(0):=0$.  The
trivial multiplicative character is denoted by $\chi^0$ and is defined
by $\chi^0(0)=0$ and $\chi^0(x)=1$ for all $x\neq 0$.  Using the equality
$\root{p^r-1}{\alpha \log_g(x)}\root{p^r-1}{\beta \log_g(x)} =
\root{p^r-1}{(\alpha+\beta)\log_g(x)}$, it is easy to see that the
pointwise multiplication  between two characters establishes the
isomorphism  $\hat{\F}_{p^r}^* \simeq \Z/(p^r-1)\Z$.

\begin{definition}[Gau{ss} sums over Finite Fields]
For the finite field $\F_{p^r}$, the multiplicative character $\chi$,
and the additive character $e_\beta$, we define the \emph{Gau{ss}
sum} $G$ by
\begin{eqnarray}
G(\F_{p^r},\chi,\beta) & := &
\sum_{x\in\F_{p^r}}{\chi(x)\root{p}{\Tr(\beta x)}}.
\end{eqnarray}
\end{definition}
\begin{example}
Let $\chi:\F_5\rightarrow \{0,1,-1,\i,-\i\}$ be the multiplicative
character defined by: $\chi(0)=0$, $\chi(1)=1$, $\chi(2)=\i$,
$\chi(3)=-\i$ and $\chi(4)=-1$.  We see that $G(\F_5,\chi,1) = \zeta_5
+ \i\root{5}{2} -\i \root{5}{3}-\root{5}{4} =
\smfrac{1}{4}\sqrt{10+2\sqrt{5}}(1-\sqrt{5}-2\i) =
\sqrt{5}\cdot\e^{2\pi\i \cdot 0.338\dots}$.
\end{example}

Obviously, $G(\F_{p^r},\chi^0,0) = p^r-1$, $G(\F_{p^r},\chi^0,\neq 0)
= -1$, and  $G(\F_{p^r},\chi,0) = 0$ for $\chi\neq \chi^0$.  In
general, for $\beta \neq 0$ we have the following fact.
\begin{fact}\label{ft:beta}
For $\beta\neq 0$ it holds that  $G(\F_{p^r},\chi,\beta\delta) =
\chi(\beta^{-1})G(\F_{p^r},\chi,\delta)$.
\end{fact}
\begin{proof}
For $G(\F_{p^r},\chi,\beta\delta)$ we have
$\sum_{x \in \F_{p^r}}{\chi(x)\root{p}{\Tr(\beta\delta x)}} =
\chi(\beta^{-1})\sum_{z \in \F_{p^r}}{\chi(z)\root{p}{\Tr(\delta z)}}$,
where we used the substitution $x \leftarrow z\beta^{-1}$ and the
multiplicativity of $\chi$.
\end{proof}

From now on we will assume that the Gau{ss} sum concerns a nontrivial
character $\chi$ and $\beta \neq 0$.  The inverse of a character
$\chi$ is defined by  $\chi^{-1}(x) := \overline{\chi(x)}$ for all
$x\neq 0$ and $\chi^{-1}(0):=0$ (where $\overline{z}$ is the complex
conjugate of $z$).  It is known that the norm of a Gau{ss} sum obeys
$|G(\F_{p^r},\chi,\beta)| = \sqrt{p^r}$, and more specifically
$G(\F_{p^r},\chi,\beta)G(\F_{p^r},\chi^{-1},\beta) = \chi(-1)p^r$.

\subsection{The Approximate Gau{ss} Sum Problem}
If we want to define the problem of estimating Gau{ss} sums as 
a computational task, we have to make clear what the length 
of the input is.  As stated above, any multiplicative character
$\chi:\F_{p^r}\rightarrow \C$ can be described by a triplet 
$(p^r,g,\alpha)$, where $g\in\F_{p^r}^*$ is a generator of $\F^*_{p^r}$
and $\alpha\in\F_{p^r}$ the index of $\chi$.
As a result, the specification of the problem 
``What is $G(\F_{p^r},\chi,\beta)$?'' as defined below, 
requires no more than $O(r\log p)$ bits of information.  
\begin{definition}[Gau{ss} Sum Problem for Finite Fields]
Let $\F_{p^r}$ be a finite field, $\chi$ a nontrivial character over
$\F_{p^r}$ and $\beta\in\F_{p^r}^*$. What is (approximately) the angle
$\gamma\bmod{2\pi}$ in the  Gau{ss} sum equation
$G(\F_{p^r},\chi,\beta) = \sqrt{p^r}\cdot \e^{\i\gamma}$?
\end{definition}
A \emph{quadratic character} is a nontrivial $\chi$ such that
$\chi(x)\in\{0,1,-1\}$ for all $x$.  
By the isomorphism $\hat{\F}^*_{p^r}\simeq \Z/(p^r-1)\Z$ 
one sees that such a character is only possible  if $p$ is odd 
and where $\chi$ is defined by $\chi(g^j)=(-1)^j$.
Unlike the case of general characters, the Gau{ss} sums of 
such quadratic characters  are known completely: $G(\F_{p^r},\chi,1) =
-(-1)^r\sqrt{p^r}$ if $p=1\bmod{4}$, and
$G(\F_{p^r},\chi,1)=-(-\i)^r\sqrt{p^r}$ if $p=3\bmod{4}$.
(See Theorem~11.5.4 in \cite{gausssums} for a proof.)

\section{Quantum Computing}\label{sec:quant}
In this section we give a brief overview of the known results
on quantum computation that are relevant for the rest of
this article.  For more information, we refer the reader 
to \cite{nielsenchuang}.

\subsection{Efficient Quantum Procedures}
\begin{fact}[Quantum Phase Estimation]\label{ft:phest}
Let $\gamma$ be an unknown phase $\bmod{2\pi}$ of the qubit state
$\ket{x_\gamma} :=
\smfrac{1}{\sqrt{2}}(\ket{0}+\e^{\i\gamma}\ket{1})$.  
If we measure this qubit in the orthogonal basis
$\ket{m_\phi} := \smfrac{1}{\sqrt{2}}(\ket{0}+\e^{\i\phi}\ket{1})$ 
 and 
$\ket{m^\perp_\phi} := \smfrac{1}{\sqrt{2}}(\ket{0}-\e^{\i\phi}\ket{1})$,
then the respective outcome probabilities are
$\Prob(m_\phi|x_\gamma) = \smfrac{1}{2}+\smfrac{1}{2}\cos(\gamma-\phi)$
and $\Prob(m_\phi|x_\gamma) = \smfrac{1}{2}-\smfrac{1}{2}\cos(\gamma-\phi)$.
Hence, if we can sample $t$ copies of $\ket{x_\gamma}$ (with various
different angles $\phi$), then we can obtain an estimate $\tilde{\gamma}$ 
of the unknown $\gamma$ within an expected error of $O(\smfrac{1}{t})$.
\end{fact}

Shor's famous article \cite{shor} implies the following result.
\begin{fact}[Efficient Quantum Algorithm for the Discrete Logarithm]\label{ft:dl}
There exists a quantum algorithm that, given a base $g\in(\Z/n\Z)^*$
and an element $x = g^j \bmod{n}$,  determines the discrete logarithm
$\log_g(x):=j$ in time  polylog($n$).
\end{fact}

\begin{fact}[Efficient Quantum Fourier Transform]
Let $\beta \in \F^*_{p^r}$.  The quantum Fourier transform
$\mathcal{F}_\beta$ over the finite field $\F_{p^r}$, which 
is defined  as the unitary mapping
\begin{eqnarray}
{\mathcal{F}_\beta}:\ket{x} & \longmapsto &
\frac{1}{\sqrt{p^r}}\sum_{y\in\F_{p^r}}{\root{p}{\Tr(\beta
xy)}\ket{y}}
\end{eqnarray}
for every $x\in \F_{p^r}$,
 can be implemented efficiently on a quantum computer.
Similarly, we can also perform the Fourier transform over the
group $\Z/n\Z$ in an efficient way.
\end{fact}
Sometimes we use the hat notation in $\mathcal{F}:\ket{\psi}\mapsto\ket{\hat{\psi}}$  
to denote the Fourier transform of a state.

\subsection{Quantum State Preparation}
For every function $f:S\rightarrow \C$, we define the state
\begin{eqnarray}
\ket{f} :=  \frac{1}{\|f\|_2}\sum_{x\in S}{f(x)\ket{x}} &  \text{with
the $\ell_2$ norm} &  \|f\|_2 :=  \sqrt{\sum_{x\in S}{|f(x)|^2}}.
\end{eqnarray}
We also allow ourselves to use the shorthand  $\ket{S} :=
\frac{1}{\sqrt{|S|}}\sum_{x\in S}{\ket{x}}$, for any set $S$.

In this article we are mostly concerned with the preparation 
of states $\ket{\chi}$ that refer to a multiplicative 
character $\chi:R\rightarrow\C$, which is zero for those
values that are not in the multiplicative subgroup $R^*$ 
and that are powers of $\root{|R^*|}{}$ for the elements
that are in $R^*$.

\begin{fact}[Phase Kickback Trick \cite{revisited}]\label{ft:phkbt}
If the computation $\ket{x}\mapsto\ket{x}\ket{f(x)}$ with 
$f(x)\in\Z/n\Z$ can be performed efficiently, 
then the phase changing transformation
$\ket{x}\mapsto\root{n}{f_x}\ket{x}$ can be performed exactly
and coherently in time polylog($n$) as well.
\end{fact}
\begin{proof}
First, create the state
$\ket{x}\ket{\hat{1}}  := 
\ket{x}\ox\frac{1}{\sqrt{n}}\sum_{j=0}^{n-1}{\root{n}{j}\ket{j}}$,
by applying the Fourier transform over $(\Z/n\Z)$ to the rightmost 
part of the initial state $\ket{x}\ket{1}$.  
Next, consider the evolution that is established by subtracting  
$f(x) \bmod{n}$ to that same rightmost register:
\begin{eqnarray}
\ket{x}\otimes\frac{1}{\sqrt{n}}
\sum_{j=0}^{n-1}{\root{n}{j}\ket{j}} &\longmapsto&
\ket{x}\otimes\frac{1}{n}
\sum_{j=0}^{n-1}{\root{n}{j}\ket{j-f(x)}} ~ = ~
\root{n}{f(x)}\ket{x}\otimes\frac{1}{\sqrt{n}}
\sum_{k=0}^{n-1}{\root{n}{k}\ket{k}},
\end{eqnarray}
where we used the substitution $j \leftarrow k+ f(x)$  and the
additivity $\root{n}{k+f(x)} =  \root{n}{f(x)} \cdot \root{n}{k}$. 
Clearly, the overall phase change of this transformation 
is the one desired. 
\end{proof}

The phase kickback trick enables us to induce the character values 
$\chi(x)$ as phases in a quantum state.
For those $x$ that have $\chi(x)=0$ we will use the 
amplitude amplification process of Grover's search algorithm
to change the amplitudes of the states $\ket{x}$.

\begin{fact}[Quantum Amplitude Amplification]\label{ft:stateprep}
Let $f:S\rightarrow \{0,1\}$ be function of which we know the
total `weight' $\|f\|_1 := \sum_{x\in S}{f(x)}$, 
but not the specific positions for which $f(x)=1$.
The corresponding state  $\ket{f}$ can be efficiently and 
exactly prepared on a quantum computer  with 
$O(\sqrt{|S|/\|f\|_1})$ queries to the function $f$.
\end{fact}
\begin{proof}
See the standard literature (\cite{searching,grover1,grover2} for example).
\end{proof}

The Facts~\ref{ft:phkbt} and \ref{ft:stateprep} show that it is 
easy to create the state
$\ket{\chi} := \frac{1}{\sqrt{p^r-1}}\sum_{x\in\F_{p^r}}{\chi(x)\ket{x}}$
with a constant number of queries to the function $\chi$.  
Furthermore, we know that $\chi$, specified by the triplet 
$(p^r,g,\alpha)$, is defined by $\chi(x)  =  \root{p^r-1}{\alpha \log_g(x)}$
for $x\in\F_{p^r}$ and $\chi(0)=0$. 
Using Shor's discrete logarithm algorithm (Fact~\ref{ft:dl}), 
we can calculate this discrete log, from which it follows
that given $(p^r,g,\alpha)$, we can create the  state $\ket{\chi}$
efficiently in the following way.

\begin{lemma}[Efficient $\chi$ state preparation]
For a finite field $\F_{p^r}$ and $(p^r,g,\alpha)$ the specification of a
multiplicative character $\chi$, the state
\begin{eqnarray}
\ket{\chi} & := &
\frac{1}{\sqrt{p^r-1}}\sum_{x\in\F{p^r}}{\chi(x)\ket{x}},
\end{eqnarray}
and its Fourier transform $\ket{\hat{\chi}}$ can be created in
polylog($p^r$) time steps on a quantum computer.
\end{lemma}
\begin{proof}
First, use the amplitude amplification process on the set $\F_{p^r}$
and the Fourier transform over $\Z/(p^r-1)\Z$ to
create the initial state
\begin{eqnarray}
\ket{\F^*_{p^r}}\ket{\hat{1}} & := &
\frac{1}{\sqrt{p^r(p^r-1)}}
\sum_{x\in\F^*_{p^r}}{\ket{x}\sum_{j=0}^{p^r-2}{\root{p^r-1}{j}\ket{j}}}.
\end{eqnarray}
Next, in superposition over all $x\in\F^*_{p^r}$ states,
calculate the discrete logarithm values $\log_g(x)$ 
and subtract  $\alpha\log_g(x)\bmod{(p^r-1)}$ 
to the state in the rightmost register. 
By the phase kickback trick of Fact~\ref{ft:phkbt} we thus obtain 
the desired state:
\begin{eqnarray}
\ket{\F_{p^r}^*}\ket{\hat{1}} & \longmapsto &
\frac{1}{\sqrt{p^r-1}}\sum_{x\in\F_{p^r}^*}{\root{p^r-1}{\alpha\log_g(x)}\ket{x}}\ket{\hat{1}}
~ = ~ \ket{\chi}\ket{\hat{1}}.
\end{eqnarray}
Given this construction, we can also create its Fourier transform
$\ket{\hat{\chi}}$ by using the quantum Fourier transform on
$\ket{\chi}$.
\end{proof}

\section{Estimating Gau{ss} Sums over Finite Fields}
\label{sec:alg}
With the ingredients of the last two sections, we are now 
ready to describe the quantum algorithm that estimates
the angle $\gamma$ of the Gau{ss} sum $G=|G|\cdot\e^{\i\gamma}$ 
over finite fields. The crucial part of our algorithm relies
on the interaction between the Fourier transform $\mathcal{F}_\beta$ 
and the multiplicative character $\chi$.
Using the fact that for nontrivial characters
$\hat{\chi}=G(\F_{p^r},\chi,\beta)/\sqrt{p^r}\cdot \bar{\chi}$,
we are able to perform a $\gamma$-phase change.
By sampling this unknown phase factor we can obtain an 
arbitrary precise estimation of $\gamma$ and thus of 
$G(\F_{p^r},\chi,\beta)$.
 
\begin{algorithm}
Consider a finite field $\F_{p^r}$, a nontrivial  character $\chi$
and a $\beta\in\F^*_{p^r}$.  If we apply the quantum Fourier transform
$(\mathcal{F}_\beta)$ over this  field to the state $\ket{\chi}$,
followed by a phase change $\ket{y}\mapsto \chi^{2}(y)\ket{y}$,  then
we generate an overall phase change according to
\begin{eqnarray}
\ket{\chi} := \frac{1}{\sqrt{p^r-1}}\sum_{x \in
\F_{p^r}}{\chi(x)\ket{x}} & \longmapsto &
\frac{G(\F_{p^r},\chi,\beta)}{\sqrt{p^r}}\ket{\chi}.
\end{eqnarray} 
\end{algorithm}
\begin{proof}
First, we note that the output after the Fourier transform
$\mathcal{F}_\beta$  looks like
\begin{eqnarray}
\ket{\hat{\chi}} & := &  \frac{1}{\sqrt{p^r(p^r-1)}} \sum_{y \in
\F_{p^r}}{\left({\sum_{x \in \F_{p^r}}{\chi(x)\root{p}{\Tr(\beta
xy)}}}\right)\ket{y}}.
\end{eqnarray}
The expression between the big parentheses equals
$G(\F_{p^r},\chi,\beta y)$,  which equals
$\chi(y^{-1})G(\F_{p^r},\chi,\beta)$ for $y\neq 0$ and is zero if
$y=0$.  In sum, we thus see that
\begin{eqnarray}
\ket{\hat{\chi}} & = &
\frac{G(\F_{p^r},\chi,\beta)}{\sqrt{p^r(p^r-1)}}\sum_{y \in
\F_{p^r}^*}{\chi(y^{-1})\ket{y}},
\end{eqnarray}
such that indeed after $\ket{y}\mapsto \chi^{2}(y)\ket{y}$ we have
created the eigenstate
$\ket{\chi^{2} \circ \hat{\chi}} = 
\frac{G(\F_{p^r},\chi,\beta)}{\sqrt{p^r}}\ket{\chi}$.
\end{proof}

With the above algorithm we are now able to efficiently estimate
the  angle $\gamma$ in the equation
$G(\F_{p^r},\chi,\beta) = \sqrt{p^r}\cdot \e^{\i\gamma}$.
\begin{theorem}[Quantum Algorithm for Gau{ss} Sum Estimation 
over $\F_{p^r}$]\label{thm:qalgff}
For any $\varepsilon>0$, there exists a quantum algorithm that
estimates the phase $\gamma$ in $G(\F_{p^r},\chi,\beta)  =
\sqrt{p^r}\cdot \e^{\i\gamma}$, with expected error 
$\mathsf{E}[|\gamma-\tilde{\gamma}|]< \varepsilon$.  
The time complexity of this algorithm is bounded by
$O(\smfrac{1}{\varepsilon}\cdot\mathrm{polylog}(p^r))$.
\end{theorem}
\begin{proof}
By the earlier algorithm, we know that we can induce the  phase change
$\ket{\chi} \mapsto \e^{\i \gamma}\ket{\chi}$ in polylog($p^r$) time.
If we do this in superposition with a `stale'  component
$\varnothing$, then we have produced the  relative phase shift
$\smfrac{1}{\sqrt{2}}(\ket{\varnothing}+\ket{\chi})  \mapsto
\smfrac{1}{\sqrt{2}}(\ket{\varnothing}+ \e^{\i \gamma}\ket{\chi})$.
As described in Fact~\ref{ft:phest}, we can estimate this phase
by measuring the states along the axis  $\ket{m_\phi} :=
\smfrac{1}{\sqrt{2}}(\ket{\varnothing}+\e^{\i \phi}\ket{\chi})$ 
for different $\phi$.
After $O(\smfrac{1}{\varepsilon})$ of such observations, 
the estimate  $\tilde{\gamma}$ of the true $\gamma$ will
have expected error $\mathsf{E}[|\gamma-\tilde{\gamma}|] < \varepsilon$.
\end{proof}

\subsection{Estimation of Jacobi Sums over Finite Fields}
Closely related to Gauss sums are the \emph{Jacobi sums,}
which play an especially important role in primality testing \cite{cohen}.
\begin{definition}[Jacobi Sums over Finite Fields]
For a finite field $\F_{p^r}$ and two multiplicative characters 
$\chi$ and $\psi$, 
the  \emph{Jacobi sum} $J(\chi,\psi)$ is defined by 
\begin{eqnarray}
J(\chi,\psi)&:=& \sum_{x\in\F_{p^r}}{\chi(x)\psi(1-x)}.
\end{eqnarray}
\end{definition}

Clearly, $J(\chi,\psi)=J(\psi,chi)$.
With $\chi^0$ the trivial character and $\psi$ a nontrivial character 
we have $J(\chi^0,\chi^0)=p^r-2$, $J(\psi,\psi^{-1})=-\psi(-1)$, 
and $J(\chi^0,\psi)=-1$.  (Note that we use the convention $\chi^0(0)=0$ 
for the primitive character, not $\chi^0(0)=1$.) 
The other, less trivial, cases have the following connection 
with Gauss sums, which is proven in Section~2 of \cite{gausssums}.
\begin{fact}
For $\chi$ and $\psi$ be nontrivial multiplicative characters over 
$\F_{p^r}$, with $\chi\psi$ nontrivial as well, 
it holds that $J(\chi,\psi)=G(\F_{p^r},\chi,1)G(\F_{p^r},\psi,1)/G(\F_{p^r},\chi\psi,1)$.
As a result, $J(\chi,\psi)=\e^{\i\lambda}\cdot\sqrt{p^r}$.
\end{fact}

\begin{corollary}[Quantum Algorithm for Jacobi Sum Estimation]
Using the Gauss sum estimation algorithm of Theorem~\ref{thm:qalgff},
there exists a quantum algorithm that estimates the 
angle $\lambda\bmod{2\pi}$ in $J(\chi,\psi)=\e^{\i\cdot \lambda}\cdot\sqrt{p^r}$
with expected error $\varepsilon$ with time complexity 
$O(\smfrac{1}{\varepsilon}\cdot\mathrm{polylog}(p^r))$.
\end{corollary}

\section{The Classical Complexity of Approximating Gau{ss} Sums}\label{sec:class}
The obvious next question now is: How difficult it is to estimate
Gau{ss} sums with classical computers?  
Although we are not able to prove that this is hard, 
we can give the following reduction, which indicates
that a classical polynomial time algorithm is unlikely.

\subsection{Reducing the Discrete Log Problem to Gau{ss} Sum Estimation}
\begin{lemma}[Reduction from Discrete Log to Gau{ss} Sum Estimation]
Let $\F_{p^r}$ be a finite field with primitive element $g$, 
$\chi(g^j):=\root{p^r-1}{j}$ a multiplicative character and
$x$ an element of $\F^*_{p^r}$.  
With an oracle that $\varepsilon$-approximates the
angle $\gamma$  of the Gau{ss} sum $G(\F_{p^r},\chi,\beta)$ for
arbitrary $\beta$, we can efficiently determine, classically, the
discrete $\log_g(x)$.
\end{lemma}
\begin{proof}
With $x=g^\ell$, we try to determine this $0\leq \ell \leq p^r-2$.
For $k=1,2,3,\dots$ we observe, using Lemma~\ref{ft:beta}, that:
$G(\F_{p^r},\chi,x^k)/G(\F_{p^r},\chi,1)  = \chi(g^{-k\ell}) =
\e^{-2\pi\i k\ell/(p^r-1)}$; call this angle $\gamma_k := -2\pi
k\ell/(p^r-1)$.  Using the `powering algorithm' ($x\mapsto x^2\mapsto
x^4\cdots$ et cetera)  we can calculate $x^k$ for any $0\leq k \leq
p^r-2$ in polylog($p^r$) time,  hence we can use our oracle to
$\varepsilon$-approximate $\gamma_k$ for any such $k$.  Via the
equality $-\smfrac{\gamma_k}{2\pi}(p^r-1) = k\ell \bmod{(p^r-1)}$ this
will give us information on the value of $\ell \bmod{(p^r-1)}$ depending on $k$.  
By estimating $\gamma_k$ for $k=1,2,4,8,\dots,\approx p^r$,
we can thus get a reliable estimation of all $\log(p^r)$ bits of
$\ell$, thereby calculating the desired value $\log_g(x)=\ell$.
\end{proof}

\subsection{Galois Automorphisms and Other Homomorphisms of 
$\boldsymbol{\Q(\root{p^r-1}{},\root{p^r}{})}$}
The previous lemma shows that it is not trivial to  estimate the
Gau{ss} sum $G(\F_{p^r},\chi,\beta)$ even if we already know the value
$G(\F_{p^r},\chi,1)$.   A similar result seems to hold for the estimation of
$G(\F_{p^r},\chi^{\alpha},\beta)$ while having information on  
$G(\F_{p^r},\chi,\beta)$.

As noted earlier, $G(\F_{p^r},\chi,\beta)$ is an element of
$\Q(\root{p^r-1}{},\root{p}{})$.  Compare now the two expressions for
$G(\F_{p^r},\chi,\beta)$ and  $G(\F_{p^r},\chi^\alpha,\beta)$,
respectively,
$\sum_{j=0}^{p^r-2}{\root{p^r-1}{j}\root{p}{\Tr(\beta g^j)}}$ and 
$\sum_{j=0}^{p^r-2}{\root{p^r-1}{\alpha j}\root{p}{\Tr(\beta g^j)}}$.
This shows that under the homomorphism $\sigma_\alpha:
\Q(\root{p^r-1}{},\root{p}{}) \rightarrow
\Q(\root{p^r-1}{\alpha},\root{p}{})$, with
$\sigma_\alpha:\root{p^r-1}{}\mapsto \root{p^r-1}{\alpha}$,  we have
$\sigma_\alpha: G(\F_{p^r},\chi,\beta) \mapsto G(\F_{p^r},\chi^\alpha,\beta)$.
(If $\gcd(\alpha,p^r-1)=1$ then this mapping is a Galois automorphism
$\sigma_\alpha \in \Gal(\Q(\root{p^r-1}{},\root{p}{})/\Q(\root{p}{}))$.  If
$\gcd(\alpha,p^r-1)\neq 1$ then the mapping $\sigma_\alpha$ is not  necessarily
bijective, and hence not an automorphism.)

This result suggests that knowledge about the Gau{ss} sum
$G(\F_{p^r},\chi,\beta)$ is sufficient to efficiently determine
$G(\F_{p^r},\chi^\alpha,\beta)$ for all other $\alpha$.  However, it should be
noted that the degree $[\Q(\root{p^r-1}{},\root{p}{}):\Q(\root{p}{})]$ equals 
$\phi(p^r-1)$, which is  exponential in the input size $\log(p^r)$. As
a result, the $\sigma_\alpha$  mapping concerns an exponential number of
coefficients, and is hence not efficient.

\subsection{Gau{ss} Sums as Pseudorandom Walks in $\boldsymbol{\C}$}
Let the finite field be a base field $\F_p$.
For every $x\neq 0$, the terms $\chi(x)e_\beta(x)$ in the summation
$\sum_x{\chi(x)e_\beta(x)}$ are unit norm vectors  in $\C$ that 
together describe a walk in $\C$ (of $p-1$ steps) from $0$ to the 
final outcome $G(\F_{p},\chi,\beta)$.
Viewed like this, an obvious classical attempt to approximate $G$
consists of trying to estimate the `average direction' of the terms 
${\chi(x)e_\beta(x)}$ by sampling a small number of $x$ values.
It should also be obvious that this method will not work for random 
samples that are not polynomial in $p$. 
As the final destination $G$ is only $\sqrt{p}$ away from the origin, 
a significant average direction can only be obtained with a sample
size that is polynomial in $p$. 

\begin{figure}
\psfrag{0}{$0$} \psfrag{30}{$30$} \psfrag{60}{$60$} \psfrag{90}{$90$}
\psfrag{120}{$120$} \psfrag{150}{$150$} \psfrag{180}{$180$}
\psfrag{210}{$210$} \psfrag{240}{$240$} \psfrag{270}{$270$}
\psfrag{300}{$300$} \psfrag{330}{$330$} \psfrag{  10}{$10$ } 
\psfrag{  20}{  $20$} \psfrag{Pseudo Random Walk 1}{$R_{t+1} := R_t +
\chi(t+1)e(t+1)$} \psfrag{Pseudo Random Walk 2}{$R'_{t+1} := R'_t +
\chi(g^t)e(g^t)$}
\begin{center}
\fbox{\scalebox{0.8}{\includegraphics{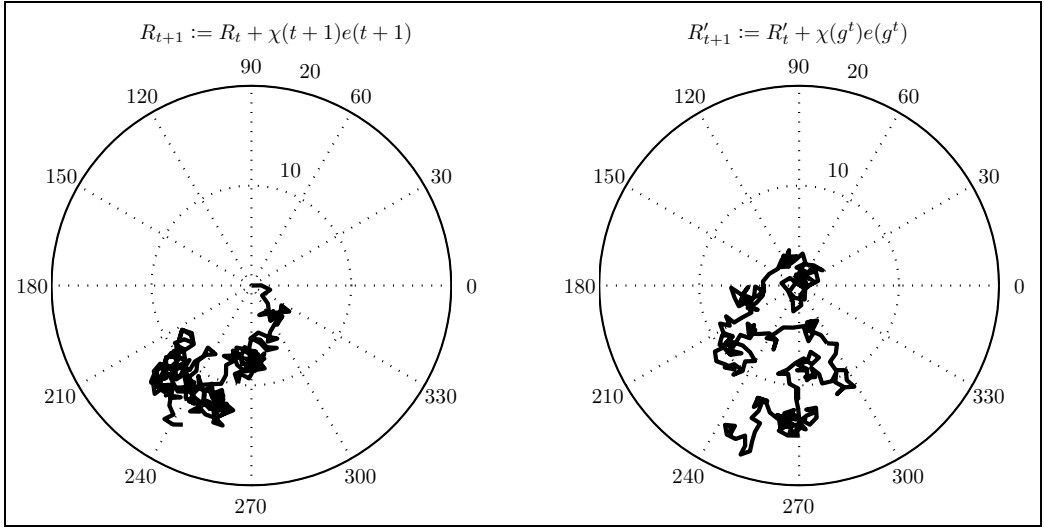}}}
\end{center}
\caption{Illustration of the pseudorandom walks described in Example~\ref{ex:prwalk}.
The walk $R$ on the left is defined by the equation 
$R(t):=\sum_{x=0}^{t}{\chi(t)e(t)}$, while the walk $R'$ 
on the right obeys $R'(t):=\sum_{j=0}^{t-1}{\chi(7^t)e(7^t)}$.
\label{fig:prwalk}}
\end{figure}

In fact, the just described walk shares many of the properties that 
a truly random walk in $\C$ would have.  
Not only does the final distance coincide with the expected 
distance norm of a random walk, but also the sequence
of steps exhibits the nonregularity of a random process.  
It is easy to verify that the autocorrelation of 
the sequence $\chi(1)e(1),\chi(2)e(2),\dots$  
is near-zero: $\mathsf{E}[\chi(j)e(j){\bar{\chi}(j+s)\bar{e}(j+s)}]=\smfrac{-e(-s)}{p-1}$ 
for $s\neq 0$.
These pseudorandom characteristics do not change when we indexing of 
the summation (and hence of the sequence) to 
$\chi(1)e(1),\chi(g)e(g),\chi(g^2)e(g^2),\dots$ with $g$ a generator of
$\F^*_p$, as this sequence obeys
$\mathsf{E}[\chi(g^j)e(g^j){\bar{\chi}(g^{j+s})\bar{e}(g^{j+s})}]=\smfrac{-\chi(-s)}{p-1}$.
See the following example for an illustration of this pseudorandom
behavior.
\begin{example}\label{ex:prwalk}
Consider the Gau{ss} sum for the finite field $\F_{241}$, with
multiplicative generator $7$, and the character defined  by $\chi(7^j)
:= \root{240}{10 j}$.    Calculations show that $G(\F_{241},\chi,1) =
\sqrt{241}\cdot \e^{2\pi\i\cdot 0.6772\dots} \approx -6.85 + 13.9\i$.
Figure~\ref{fig:prwalk} shows the two pseudorandom walks that 
are defined by the sequences $\chi(1)e(1),\chi(2)e(2),\dots$ (left) 
and $\chi(7^0)e(7^0),\chi(7^1)e(7^1),\dots$ (right).
\end{example}

\section{Gau{ss} Sums over Finite Rings}\label{sec:rings}

Although the final quantum algorithm for estimating Gau{ss} sums
over rings $\Z/n\Z$ is not much more complicated than the
finite field algorithm, the theory surrounding it is somewhat
more elaborate.  A large part of this section concerns 
the proper description of a multiplicative character over $\Z/n\Z$
and its various properties.  
These details are necessary to get a valid definition 
for the input size of the Gau{ss} sum problem over $\Z/n\Z$
(see Definition~\ref{def:dir}).

\subsection{Definitions and Notation: Dirichlet Characters}\label{sec:defs2}
Again, the $n$th root of unity is denoted by $\root{n}{x}:= \e^{2\pi\i
x/n}$.  From \cite{cohen}, Section~1.4 we copy the following facts.
Consider the multiplicative subgroup $(\Z/n\Z)^*$ with the prime
decomposition $n=p_1^{r_1}\cdots p_k^{r_k}$.  Following the Chinese
remainder theorem we have $(\Z/n\Z)^* \simeq
(\Z/p_1^{r_1}\Z)^*\times\cdots \times(\Z/p_k^{r_k}\Z)^*$,  such that
$|(\Z/n\Z)^*| = \phi(n)$, with $\phi$ Euler's totient function.
Furthermore we have
\begin{eqnarray}\label{eq:equiv}
\left\{\begin{array}{rcl} (\Z/p^r\Z)^* & \simeq & \Z/(p-1)p^{r-1}\Z
\text{ if $p\geq 3$}, \\ (\Z/2\Z)^* & \simeq & \Z/\Z~ \simeq~ \{0\},\\
(\Z/4\Z)^* & \simeq & \Z/2\Z, \\ (\Z/2^r\Z)^* & \simeq & \Z/2\Z \times
\Z/2^{r-2}\Z \text{ if $r\geq 3$,}
\end{array}\right. & & 
\end{eqnarray}
hence $(\Z/n\Z)^*$ is cyclic if and only if $n=2,4,p^r$ or $2p^r$,
with $p$ an odd prime.

\begin{definition}[Dirichlet Characters]
A function $\chi:\Z/n\Z\rightarrow \C$ is a Dirichlet  character if
for all $x,y\in\Z/n\Z$ we have $\chi(x)\chi(y)=\chi(xy)$ and
$\chi(x)=0$ if and only if $\gcd(n,x)\neq 1$.
\end{definition}
Using the multiplicative decomposition  $(\Z/n\Z)^* =
(\Z/p_1^{r_1}\Z)^*\times \cdots \times (\Z/p_k^{r_k}\Z)^*$,  we see that 
all Dirichlet characters $\chi:(\Z/n\Z)\rightarrow \C$ can  be decomposed
as $\chi=\chi_1\cdots\chi_k$ with $\chi_i:(\Z/p_i^{r_i}\Z)\rightarrow\C$ 
for every $1\leq i \leq k$, and thus 
$\chi(x):= \chi_1(x\bmod{p_1^{r_1}})\cdots\chi_k(x\bmod{p_k^{r_k}})$.

For $p$ an odd prime the character $\chi:\Z/p^r\Z\rightarrow \C$ can
be  described by the expression $\chi(x) := \root{\phi(p^r)}{\alpha
\log_g(x)}$, where $g$ is a generator of the cyclic $(\Z/p^r\Z)^*$,
$\phi(p^r) = p^{r-1}(p-1)$ and $\alpha\in\Z/\phi(p^r)\Z$.  For
$(\Z/2\Z)^*$ we only have the trivial character $\chi^0$, while for
$(\Z/4\Z)$ we have two possibilities: $\chi^0$ and $\chi^1$ with
$\chi^{\alpha}(3)=(-1)^\alpha$ and $\chi^{\alpha}(1)=1$.  If $\chi$ is
a character over $(\Z/2^r\Z)^*$ with $r\geq 3$,  then we have to
decompose the character in two terms.  The group $(\Z/2^r\Z)^*$ is
generated by $3$ and $5$ (see \cite{cohen}), hence the character can
be described by the pair $(\alpha,\alpha')\in(\Z/2\Z)\times(\Z/2^{r-2}\Z)$
such that for all $i$ and $i'$ we have 
$\chi(3^i5^{i'} \bmod{2^r}) := 
(-1)^{\alpha i}\root{2^{r-2}}{\alpha' i'}$ (while $\chi(x)=0$ if 
$x$ is even).

\begin{definition}[Specification of Dirichlet Characters]\label{def:dir}
Let  $(\Z/n\Z)^* \simeq
(\Z/2^{r_0}\Z)^*\times(\Z/p_1^{r_1}\Z)^*\times \cdots \times
(\Z/p_k^{r_k}\Z)^* \simeq  (\Z/2^{r_0}\Z)^*\times(\Z/\phi_1\Z)\times
\cdots \times (\Z/\phi_k\Z)$, with $\phi_j := (p_j-1)p_j^{r_j-1}$ (see
Equation~\ref{eq:equiv}).  The specification of a Dirichlet character
$\chi:\Z/n\Z\rightarrow \C$  is done by three sequences $(p,g,\alpha)$,
with the prime decomposition $p=(p_1,\dots,p_k)$ of $n$, the generators
$g=(g_1,\dots,g_k)\in(\Z/p_1^{r_1}\Z)^*\times\cdots\times(\Z/p_k^{r_k}\Z)^*$
of the multiplicative groups, and
$\alpha=((\alpha_0,\alpha'_0),\alpha_1,\dots,\alpha_k)\in  (\Z/2\Z
\times \Z/2^{r-2})\times(\Z/\phi_1\Z)\times\cdots\times(\Z/\phi_k\Z)$
the specification of the characters $\chi_j$ in the definition
$\chi(x):=\chi_0(x\bmod{2^{r_0}})\chi_1(x\bmod{p_1^{r_1}})\cdots\chi_k(x\bmod{p_k^{r_k}})$
with
\begin{eqnarray}
\chi_0(3^i5^{i'}\bmod{2^{r_0}}) := 
(-1)^{\alpha_0 i}\root{2^{r_0-2}}{\alpha'_0i'} & \text{and} &
\chi_j(x_j) := \root{\phi_j}{\alpha_j \log_{g_j}(x_j)} \text{if
$x_j\in(\Z/p_j^{r_j}\Z)^*$},
\end{eqnarray}
while $\chi_j(x_j):=0$ if $\gcd(x,p_j)\neq 1$.
\end{definition}
With this definition, we see that the specification of 
a Dirichlet character $\chi:\Z/n\Z\rightarrow \C$
requires only $O(\log n)$ bits of information. 
Hence, in the context of such characters, an algorithm 
that requires polylog($n$) steps is efficient, while a 
running time polynomial in $n$ is inefficient.

\begin{lemma}[Calculation of Dirichlet Character Values] 
Let $(p,g,\alpha)$ be the specification of a character
$\chi:\Z/n\Z\rightarrow \C$.  Given $n$ and $(p,g,\alpha)$, we can
induce the phase change  $\ket{x}\mapsto \chi(x)\ket{x}$ for any
$x\in(\Z/n\Z)^*$ efficiently with a quantum algorithm.
\end{lemma}
\begin{proof}
To compute the phase change $\ket{x}\mapsto \chi_j(x\bmod{p_j^{r_j}})\ket{x}$ 
we perform the following two steps (we use a similar protocol 
for the $\chi_0$ part of $\chi$):
\begin{enumerate}
\item Use Shor's discrete log algorithm (Fact~\ref{ft:dl}) 
to determine the value
$s_j:=\log_{g_j}(x\bmod{p_j^{r_j}})$.
\item Use the phase kickback trick (Fact~\ref{ft:phkbt})
to induce the phase change
$\ket{x}\mapsto\root{\phi(p_j^{r_j})}{\alpha_j s_j}$.
\end{enumerate} 
Perform the phase changes for all $\chi_j$ to the same $x$ state,  
such that the overall transformation will be: $\ket{x}\mapsto \chi_0(x)\ket{x} \mapsto
\chi_0(x)\chi_1(x)\mapsto \cdots \mapsto
\chi_{0}(x)\chi_1(x)\cdots\chi_k(x)\ket{x} = \chi(x)\ket{x}$.
\end{proof}

\begin{definition}[Conductance and Triviality of Dirichlet Characters]
A character $\chi$ is \emph{trivial} if $\chi(x)=1$ for all
$x\in(\Z/n\Z)^*$, that is, if $\alpha$ is the zero vector.
The \emph{conductor} $c$ of a character
$\chi:\Z/n\Z\rightarrow \C$ is the minimum value $c>1$ for which there
is a character  $\chi_c:\Z/c\Z\rightarrow \C$ such that
$\chi=\chi_c\chi^0$ where $\chi^0$ is the trivial character over
$\Z/n\Z$.  We call $\chi$ a \emph{primitive} character if  $\chi$ has
the maximum conductance $c=n$.  The trivial character has conductance
$1$ and is not primitive.  The field $\Z/p\Z$ has $p-2$ primitive
characters,  whereas the ring $\Z/p^r\Z$ with $r\geq 2$ has
$p^{r-2}(p-1)^2$ such characters.   For $\Z/2^r\Z$ with $r\geq 2$ this
means that all characters $\chi^{(1,\alpha')}$ are primitive ($\Z/2\Z$
has no primitive characters, and $\Z/4\Z$ has one).  For $\Z/p^r\Z$
with $p$ and odd prime, a character  $\chi(x) =
\root{\phi(p^r)}{\alpha\log_g(x)}$  has conductance $p^{r-s}$ where
$p^s\mid\alpha$, while $p^{s+1}\nmid \alpha$. 
As a result $\chi^\alpha$ is primitive, if and only if $p\nmid
\alpha$.  In general, a character over the ring $\Z/n\Z$ with
$(\Z/n\Z)^*\simeq (\Z/p_1^{r_1}\Z)^*\times \cdots \times
(\Z/p_k^{r_k}\Z)^*$, and accordingly $\chi=\chi_1\cdots\chi_k$, is
primitive if and only if all $\chi_i$ factors are primitive.
\end{definition}

\subsection{Definition and Properties of Gau{ss} Sums over Rings $\boldsymbol{\Z/n\Z}$}
The definition of Gau{ss} sums over $\Z/n\Z$ is a natural 
 generalization of the  definition for finite fields.
\begin{definition}[Gau{ss} sums over Finite Rings]
For the ring $\Z/n\Z$, the Dirichlet character $\chi$,  and the
additive character $e_\beta(x):=\root{n}{x\beta}$,  we define the
\emph{Gau{ss} sum} $G$ by
\begin{eqnarray}
G(\Z/n\Z,\chi,\beta) & := &  \sum_{x\in\Z/n\Z}{\chi(x)\root{n}{\beta
x}}.
\end{eqnarray}
\end{definition}
Note that the Gau{ss} sum terms are the coefficients of the Fourier
transform of the character: $G(\Z/n\Z,\chi,\beta) =
\hat{\chi}(\beta)$, with the Fourier transform over the additive group
$\Z/n\Z$.
See Section~1.6 in \cite{gausssums} for the proofs of the facts below.

\begin{fact}\label{ft:nontrivzerosum}
\newcounter{cr:nontrivzerosum}
\setcounter{cr:nontrivzerosum}{\value{fact}} Let $\chi$ be a
nontrivial Dirichlet character over $\Z/n\Z$.   The summation of all
$\chi$ values obeys
$\sum_{x=0}^{n-1}{\chi(x)} = 0$.
If $\chi$ is trivial, then the sum equals $\phi(n)$.
\end{fact}

\begin{fact}\label{ft:charfactor}
\newcounter{cr:charfactor} \setcounter{cr:charfactor}{\value{fact}}
Let $\chi$ be a character over the ring  $(\Z/n\Z)^* \simeq
(\Z/p_0^{r_0}\Z)^*\times\cdots \times(\Z/p_k^{r_k}\Z)^*$,  such that
$\chi = \chi_0\cdots \chi_k$ with $\chi_i$ a multiplicative character
over $(\Z/p_i^{r_i}\Z)^*$ for every $0 \leq i\leq k$.  With $J_i \in
(\Z/p_i^{r_i}\Z)^*$ the integers such that  $J_i n/p_i^{r_i} = 1
\bmod{p_i^{r_i}}$, it holds that
$G(\Z/n\Z,\chi,\beta) = 
\prod_{i=0}^k{G(\Z/p_i^{r_i}\Z,\chi_i,\beta J_i)}$.
(Such $J_i$ always exist because $\gcd(p_i,n/p_i^{r_i})=1$, and  hence
$n/p_i^{r_i} \in (\Z/p_i^{r_i}\Z)^*$, for all $i$.)
\end{fact}
This last lemma shows that we should only be concerned about Gau{ss} sums
over rings $\Z/p^r\Z$, the size of a prime power. For trivial
characters, such Gau{ss} sums are easily calculated.

\begin{fact}\label{ft:trivialchar}
\newcounter{cr:trivialchar} \setcounter{cr:trivialchar}{\value{fact}}
Let $\chi^0:\Z/p^r\Z \rightarrow \C$ be the trivial character, and
$p^j\mid\beta$ with $p^{j+1}\nmid \beta$, then
\begin{eqnarray}
G(\Z/p^r\Z,\chi^0,\beta) & = &  \left\{
\begin{array}{rl}
p^{r-1}(p-1) & \text{if $j=r$,} \\ -p^{r-1} & \text{if $j=r-1$,} \\ 0
& \text{if $j< r-1$}.
\end{array}
\right.
\end{eqnarray} 
\end{fact} 
Nontrivial characters that are not primitive can be reduced to 
primitive characters over smaller groups in the following way.
\begin{fact}\label{ft:periodchar}
\newcounter{cr:periodchar} \setcounter{cr:periodchar}{\value{fact}}
Let $\chi:\Z/p^r\Z \rightarrow \C$ be a non-primitive  character with
conductance $p^{r-s}$, then the corresponding  Gau{ss} sum obeys
(note that $\chi$ modulo $p^{r-s}$ will be primitive):
\begin{eqnarray}
G(\Z/p^r\Z,\chi,\beta) & = &  \left\{
\begin{array}{rl}
p^{s-1}(p-1) \cdot G(\Z/p^{r-s}\Z,\chi,\beta/p^s) &  \text{if $\beta\mid
p^s$,}\\ 0 & \text{if $\beta\nmid p^s$.}
\end{array}
\right.
\end{eqnarray}
\end{fact}
Similar to the finite field case, the $\beta$ index of the additive character
can be `factored out' as $\chi^{-1}(\beta)$:
\begin{fact}\label{ft:primchar}
\newcounter{cr:primchar} \setcounter{cr:primchar}{\value{fact}} Let
$\chi$ be a primitive character over $\Z/n\Z$, then
$\hat{\chi}(\beta)=\chi^{-1}(\beta)\hat{\chi}(1)$, and hence,
equivalently, $G(\Z/n\Z,\chi,\beta) = \chi^{-1}(\beta)G(\Z/n\Z,\chi,1)$.
Also, $|G(\Z/n\Z,\chi,1)|=\sqrt{n}$ holds.
\end{fact}

\section{Estimating Gau{ss} Sums over Finite Rings}\label{sec:alg_ring}
\begin{theorem}[Quantum Algorithm for Gau{ss} Sum Estimation over $\Z/n\Z$]
For any $\varepsilon>0$, there exists a quantum algorithm that
 estimates the phase $\gamma$ in $G(\Z/n\Z,\chi,\beta)  =
|G(\Z/n\Z,\chi,\beta)| \cdot \e^{\i\gamma}$, with expected error
$\mathsf{E}[|\gamma-\tilde{\gamma}|]<\varepsilon$.
The time complexity of this algorithm is bounded by
$O(\smfrac{1}{\varepsilon}\cdot\mathrm{polylog}(n))$.  
Also the norm $|G(\Z/n\Z,\chi,\beta)|$ can be determined  
in polylog($n$) time.
\end{theorem}
\begin{proof}
\begin{enumerate}
\item Determine the integers $J_0,\dots,J_k$ as mentioned in
Fact~\ref{ft:charfactor}.
\item Calculate the Gau{ss} sums $G(\Z/p_i^{r_i}\Z,\chi_i,\beta J_i)$
for  the trivial characters $\chi_i$, using Fact~\ref{ft:trivialchar}.
Continue with the reduced product of nontrivial characters.
\item Re-express the Gau{ss} sums terms for the periodic characters,
using Fact~\ref{ft:periodchar}.   
Continue with the reduced product of primitive characters.
\item  Use Algorithm~\ref{alg:primchar} to 
calculate the norms and estimate the phases of the
Gau{ss} sums of the remaining primitive characters
(using the standard phase estimation technique of Fact~\ref{ft:phest}
this requires $O(\smfrac{1}{\varepsilon}\cdot\mathrm{polylog}(n))$ steps).
\end{enumerate}
\begin{algorithm}\label{alg:primchar}
Let $\chi:\Z/n\Z\rightarrow \C$ be a primitive character and
$\beta\in\Z/n\Z$.  The following algorithm 
calculates the norm and estimates the phase $\gamma$
in the Gau{ss} sum 
$G(\Z/n\Z,\chi,\beta)=|G(\Z/n\Z,\chi,\beta)|\cdot \e^{\i\gamma}$.
\begin{enumerate}
\item If $\beta\notin(\Z/n\Z)^*$ then conclude that
$G(\Z/n\Z,\chi,\beta)=0$  (see Fact~\ref{ft:primchar}).
\item Otherwise, use Shor's discrete log algorithm to determine the
$\chi(\beta^{-1})$ factor in the equality  $G(\Z/n\Z,\chi,\beta) =
\chi(\beta^{-1})G(\Z/n\Z,\chi,1)$ (Fact~\ref{ft:primchar}).  Continue
with the estimation of $\gamma$ in  $G(\Z/n\Z,\chi,1)=\e^{\i
\gamma}\sqrt{n}$.
\item Apply the quantum Fourier transform $\mathcal{F}$ over the ring
$\Z/n\Z$ to the state $\ket{\chi}$, followed by a phase change
$\ket{y}\mapsto\chi^2(y)\ket{y}$ for all $y\in(\Z/n\Z)^*$ in the
superposition $\ket{\hat{\chi}} = \sum_y{\hat{\chi}(y)\ket{y}}$. 
Because for primitive characters $\hat{\chi}(y)=\chi^{-1}(y)\hat{\chi}(1)$
(Fact~\ref{ft:primchar}),
this transformation generate an overall phase change according to
$\ket{\chi} \mapsto \e^{\i\gamma}\ket{\chi}$.
Like in Theorem~\ref{thm:qalgff}, 
we use this phase change to estimate $\gamma$ to the required precision
$\varepsilon$.
\end{enumerate} 
\end{algorithm}
\end{proof}

It should be noted that for primitive characters over rings 
$\Z/p^r\Z$ with $r\geq 2$ the Gauss sums $G(\Z/p^r\Z,\chi,1)$ 
are known (Section~1.6 in \cite{gausssums}). Hence step~3 of
the above algorithm is not always necessary. 

\section{Conclusion and Discussion}\label{sec:concl}
The algorithms that we presented in this article rely on the specific
interaction between multiplicative characters and Fourier transformations,
some of which have been described earlier in \cite{shiftquadchar} 
and \cite{hiddencoset1}.  Typical for these results is the fact 
they are defined for finite fields or finite rings but not for groups,
which indicates a departure from the Hidden Subgroup framework
for quantum algorithms \cite{hsg}.
What is new about the results presented here, is that they describe  
quantum algorithms for a natural problem that does not assume 
the presence of a black box function.
The only other natural problems for which an efficient quantum
algorithm has been constructed are those described by Shor 
\cite{shor} and Hallgren \cite{pell}, both of which deal with 
number theory as well.  

For the results of this article
it remains therefore an important open question if  Gau{ss} sum 
estimation is hard classically, even under the assumption that 
factoring and discrete logarithms are easy.  
If this is indeed the case, another related question remains: 
Which problems reduce to Gau{ss} sum estimation
that do not reduce to factoring or the discrete logarithm problems?

\section*{Acknowledgments}
We would like to thank Vinay Deolalikar,  Hendrik Lenstra and Ronny
Roth for helpful discussions about the topics in this article.

\end{document}